\documentclass{aa}
\usepackage{graphicx}
\usepackage{natbib}

\newcommand{\Msun}{M_{\odot}}
\newcommand{\pc}{{\rm \, pc}}
\newcommand{\kpc}{{\rm \, kpc}}
\newcommand{\kmps}{{\rm \, km \, s^{-1}}}
\newcommand{\kmpspkpc}{{\rm \, km \, s^{-1} \, kpc^{-1}}}

\newcommand{\K}{{\rm \, K}}

\newcommand{\myr}{\rm \,Myr}
\newcommand{\omgb}{\Omega_{\rm b}}
\newcommand{\vobsr}{V_{\rm obs}(R)}
\newcommand{\vpotr}{V_{\rm pot}(R)}
\newcommand{\vobs}{V_{\rm obs}}
\newcommand{\vpot}{V_{\rm pot}}
\newcommand{\Mobs}{M_{\rm obs}}
\newcommand{\Mpot}{M_{\rm pot}}

\newcommand{\alpcrit}{\alpha_{\rm crit}}
\newcommand{\gmmcrit}{\gamma_{\rm crit}}
\newcommand{\avrPalp}{\tilde{P}_{\alpha}}
\newcommand{\avrPgmm}{\tilde{P}_{\gamma}}

\begin{document}
   \title{Intrinsic errors of the central galactic mass derived from rotation curves
          under the influence of a weak non-axisymmetric potential}

   \subtitle{}

   \author{Jin Koda\inst{1,2}\thanks{JSPS Research Fellow; 
               Present address: ALMA-J Project Office,
               National Astronomical Observatory, 2-21-1, Osawa, Mitaka,
               Tokyo, 181-8588, Japan}
          \and
           Keiichi Wada\inst{2}
          }

   \titlerunning{Intrinsic errors of the galactic mass from rotation curves}

   \authorrunning{J. Koda \& K. Wada}

   \offprints{J. Koda\\
	\email{jin.koda@nao.ac.jp}}

   \institute{Institute of Astronomy, University of Tokyo,
              2-21-1, Osawa, Mitaka, Tokyo, 181-0015, Japan\\
         \and
              National Astronomical Observatory,
              2-21-1, Osawa, Mitaka, Tokyo, 181-8588, Japan\\
             }
   \date{Received 17 July 2002 / Accepted 1 October 2002}

   \abstract{
Rotation curves are often used to estimate the mass distribution of spiral
galaxies, assuming that the circular velocities of the interstellar medium
balance with the galactic centrifugal force. 
However, non-circular motions caused by a non-axisymmetric gravitational
potential, such as a stellar bar, may disturb the velocity field, resulting
in errors in mass estimation, especially in the central regions of
galaxies. This is because 
the line-of-sight velocity depends on the viewing angles in
a non-axisymmetric flow.
Observing rotation curves of edge-on galaxies in time-dependent
numerical simulations from different viewing angles,
we obtain errors in the estimation of galactic mass from the rotation curves.
In the most extreme case, 
the ellipticity of gas orbits is as high as $e \sim 0.8$
in the central regions, even if the bar potential is weak.
When rotation curves are defined as the highest velocity envelope of 
position-velocity diagrams, the mass estimated from the rotation curves
is larger than the true mass by a factor of five for 15\% of the viewing
angles, and the ratio between the apparent mass and true mass is less
than six for any viewing angle.
The overestimation in mass occurs more frequently than the underestimation.
   }
   \maketitle


\section{Introduction}
Rotation curves are a major tool for
determining mass distribution in spiral galaxies.
Assuming spherical mass distribution,
the galactic mass $M$ within a galactocentric radius $R$ is
estimated by
\begin{equation}
M = R V^2 / G,
\label{eq:vir}
\end{equation}
where $V$ and $G$ stand for a circular velocity and gravitational
constant. 

Rotation curves in their outer regions are generally flat
\citep{ru80, ru82, ru85}, indicating massive dark halos
surrounding their optical disks \citep{ke87}.
In their central regions, many rotation curves rise steeply
from the centers, reaching the high velocity seen
in the outer flat rotation curves: typically $100-300\kmps$
within a central $100\pc$ radius \citep{sf96,sf99}.
These high velocities may indicate central massive cores
of about $10^9\Msun$ within a central $100\pc$
radius \citep{sf96,sf99, ts00}.

However, the gas in galactic disks does not necessarily 
show pure circular rotation, especially in the central regions.
Bar-like distortions of the stellar system
can drive non-circular (elliptical)
motions for the gas. As a result, 
the apparent rotation curves do not represent the
correct mass distribution.
For example, if the elliptical orbits were aligned by chance
with the line-of-sight, we would overestimate the mass \citep{sk99}. 
The effect of non-circular motion 
on position-velocity diagrams has been intensively studied
in theoretical calculations \citep{ba99,ab99}.

In this paper, we quantitatively study the errors in estimating the
mass from rotation curves in galaxies with a weak bar,
and calculate the probability that  the observed mass suffers from such
errors.
Even if the bar-like distortion of the gravitational potential is
very weak, the gas velocity-field can be non-axisymmetric \citep{wa94}.
In order to obtain the velocity-field of the gas in a weak bar
potential, we performed Smoothed Particle Hydrodynamics (SPH) simulations
(Sect. \ref{sec:numcal}). Using the numerical results,
we estimate the probability of overestimating the central galactic mass
(Sect. \ref{sec:calprob}).
We discuss the implications of our results in Sect. \ref{sec:disc}.


\section{Numerical Calculations}\label{sec:numcal}

\subsection{2-D SPH Methods}
We perform two-dimensional SPH calculations \citep{lc77,gm77},
and reproduce gas motions in a weak stellar bar potential.
We adopt the SPH formulation by \citet{bz90}, and use the spline kernel
\citep{ml85} with the modification for its gradient \citep{tc92}.
The correction term for viscosity \citep{ba95} is taken into account to
avoid large entropy generation in pure shear flows. The SPH smoothing
length $h$ varies in space and time, keeping the number of particles within
the radius $2h$ at an almost constant of 32 
according to the method of \citet{hk89}.
The leapfrog integrator is adopted
to update positions and velocities. We use $3\times10^4$ particles
to represent the gas disk.

\subsection{Galaxy Disk and Weak Bar Potentials}
We take the galaxy potential for a weak bar used in \citet{sd77} and \citet{wh92};
the oval potential for a barred galaxy $\Phi(R,\theta)$ is represented
in axisymmetric and bisymmetric parts by
\begin{equation}
\Phi(R,\theta) = \Phi_{0}(R) + \Phi_{b}(R) \cos 2\theta,
\label{eq:pottot}
\end{equation}
where the first term is an axisymmetric potential for a disk,
\begin{equation}
\Phi_{0}(R) = - C \frac{1}{(R^2 + a^2)^{1/2}},
\label{eq:apot}
\end{equation}
and the second term is a bisymmetric one for a weak bar,
\begin{equation}
\Phi_b = - \epsilon_0 C \frac{a R^2}{(R^2+a^2)^2}.
\end{equation}
The normalization coefficient $C = (27/4)^{1/2} a v_{\rm max}^2$ is calculated
from the core radius $a$ and maximum rotation velocity $v_{\rm max}$.
$\epsilon_0$ is a parameter to control the strength of the bar. $R$ and
$\theta$ stand for the galactocentric radius and azimuthal angle from
the bar respectively.
We are interested in the notion that noncircular motions in a bar apparently
indicate a central massive component. We thus do not consider
any massive central component in our potential, such as a bulge \citep{ab99}
or a massive black hole \citep{fw98,fh00}.

Our potential model has the benefit of being capable of analytically
investigating gaseous orbits in the bar potential \citep{wa94},
and has been well-studied in numerical simulations for the bar-driven
gas fueling into galactic centers \citep{wh92, wh95}, the gas kinematics
in the Galaxy \citep{wt94}, the spatial distribution of mass-to-light
ratio in a galaxy NGC 4321 \citep{ws98}, and the effects of a central
black hole \citep{fw98, fh00}.

\begin{figure}
   \centering
   \includegraphics{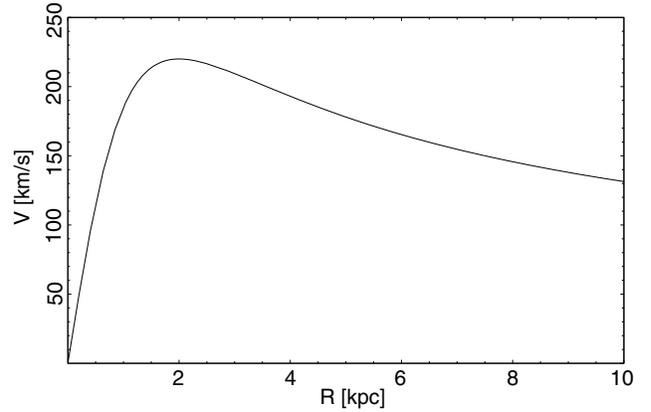}
   \caption{Rotation curve from the axisymmetric potential, eq.(\ref{eq:apot}),
    with the core radius $a=\sqrt{2}\kpc$ and
    maximum rotation velocity $v_{\rm max} = 220 \kmps$.}
              \label{fig:rotc}%
   \end{figure}

\begin{figure}
   \centering
   \includegraphics{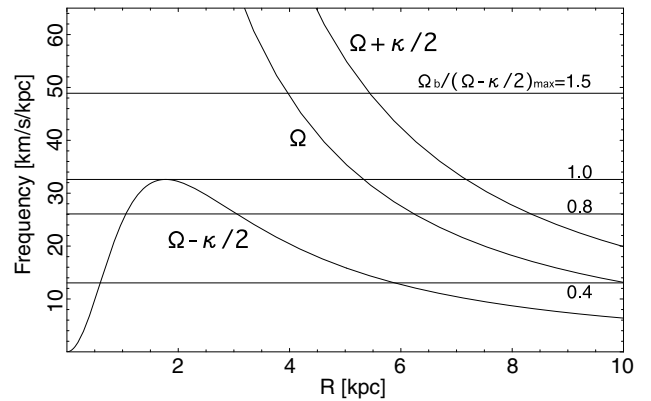}
   \caption{Radial changes of frequencies, $\Omega(R)$ and $\Omega(R) \pm \kappa(R)/2$.
    Four horizontal lines represent the pattern speeds of the bar, i.e. 0.4, 0.8, 1.0,
    and 1.5 $\times (\Omega-\kappa/2)_{\rm max}$.}
              \label{fig:freq}%
   \end{figure}

We fix $a=\sqrt{2}\kpc$ and $v_{\rm max}=220\kmps$; the corresponding rotation
curve is shown in Fig. \ref{fig:rotc}. The gas reaches the maximum circular
rotation velocity at $R=2\kpc$ with a rotational period of $56\myr$.
Fig. \ref{fig:freq} shows the radial changes of frequencies, $\Omega$ and
$\Omega\pm\kappa/2$, where $\Omega$ and $\kappa$ are circular and epicyclic
frequencies respectively. We set the pattern speed of the bar $\omgb$ at
0.4, 0.8, and 1.5 times the maximum of $\Omega-\kappa/2$, indicated by horizontal
lines. Models with $\omgb=0.4$ and $0.8\times (\Omega-\kappa/2)_{\rm max}$
have two inner Lindblad resonances (ILR),
while those with $\omgb=1.5\times (\Omega-\kappa/2)_{\rm max}$ have no ILR.
$\epsilon_0$ is set to 0.05, 0.10,
and 0.15. Our nine models are listed in Table \ref{tab:mdl}.

%

%

   \begin{table*}
      \caption[]{Model parameters}
         \label{tab:param}
     $$ 
         \begin{array}{cccccccccc}
            \hline
            \noalign{\smallskip}
 model  & \multicolumn{2}{c}{\omgb} & \epsilon_0 & \multicolumn{2}{c}{ILRs}   &    CR    &   OLR    & (\rho_b/\rho_0)_{\rm max} & \rm T_{bar}    \\
\cline{2-3} \cline{5-6}
        & {\tiny (\kmpspkpc)} & {\tiny (\Omega-\kappa/2)_{\rm max}} &          & (\kpc) & (\kpc) & (\kpc) & (\kpc) &       (\%)      &   (\myr)\\
            \noalign{\smallskip}
            \hline
            \noalign{\smallskip}
 A..... & 49 & 1.5       & 0.05         & \multicolumn{2}{c}{--} &  4.0     & 5.4      &   5             & 130\\
 B..... & 49 & 1.5       & 0.10         & \multicolumn{2}{c}{--} &  4.0     & 5.4      &  10             & 130\\
 C..... & 49 & 1.5       & 0.15         & \multicolumn{2}{c}{--} &  4.0     & 5.4      &  16             & 130\\
 D..... & 26 & 0.8       & 0.05         & 1.1 & 3.1 &  6.2     & 8.3      &   5             & 240\\
 E..... & 26 & 0.8       & 0.10         & 1.1 & 3.1 &  6.2     & 8.3      &  10             & 240\\
 F..... & 26 & 0.8       & 0.15         & 1.1 & 3.1 &  6.2     & 8.3      &  16             & 240\\
 G..... & 13 & 0.4       & 0.05         & 0.6 & 5.9 & 10.0     & 13.3     &   5             & 470\\
 H..... & 13 & 0.4       & 0.10         & 0.6 & 5.9 & 10.0     & 13.3     &  10             & 470\\
 I..... & 13 & 0.4       & 0.15         & 0.6 & 5.9 & 10.0     & 13.3     &  16             & 470\\
            \noalign{\smallskip}
            \hline
         \end{array}
     $$ 
\begin{list}{}{}
\item[] Two free parameters, i.e. bar pattern speed $\omgb$ and bar strength
$\epsilon_0$, and radii of inner Lindblad resonances (ILR), corotation
resonance (CR) and outer Lindblad resonance (OLR), maximum density ratio
of the bar over the disk $(\rho_b/\rho_0)_{\rm max}$,
and rotational time-scale of the bar
$\rm T_{bar} [\equiv 2 \pi / \omgb]$ at $R=2\kpc$ are tabulated.
\end{list}
\label{tab:mdl}
   \end{table*}

\subsection{Initial Conditions}

The gas is initially distributed in a uniform-density disk 
with an $8\kpc$ radius,
following pure circular-rotation that balances the centrifugal force.
The gas temperature is assumed to be a constant $10^4\K$, 
corresponding to the sound
speed of about $10\kmps$, throughout evolution. The total gas mass is assumed to
be 5\% of the total stellar mass within the radius of $8\kpc$.
The results are not significantly affected by the total gas mass,
because thermal pressure is much smaller than the rotational energy, 
and we do not calculate self-gravity of the gas.
We advance the calculations up to about $500\myr$.

\subsection{Gas Dynamical Evolution}\label{sec:gde}

%

   \begin{figure*}
   \centering
   \includegraphics[width=0.75\textwidth]{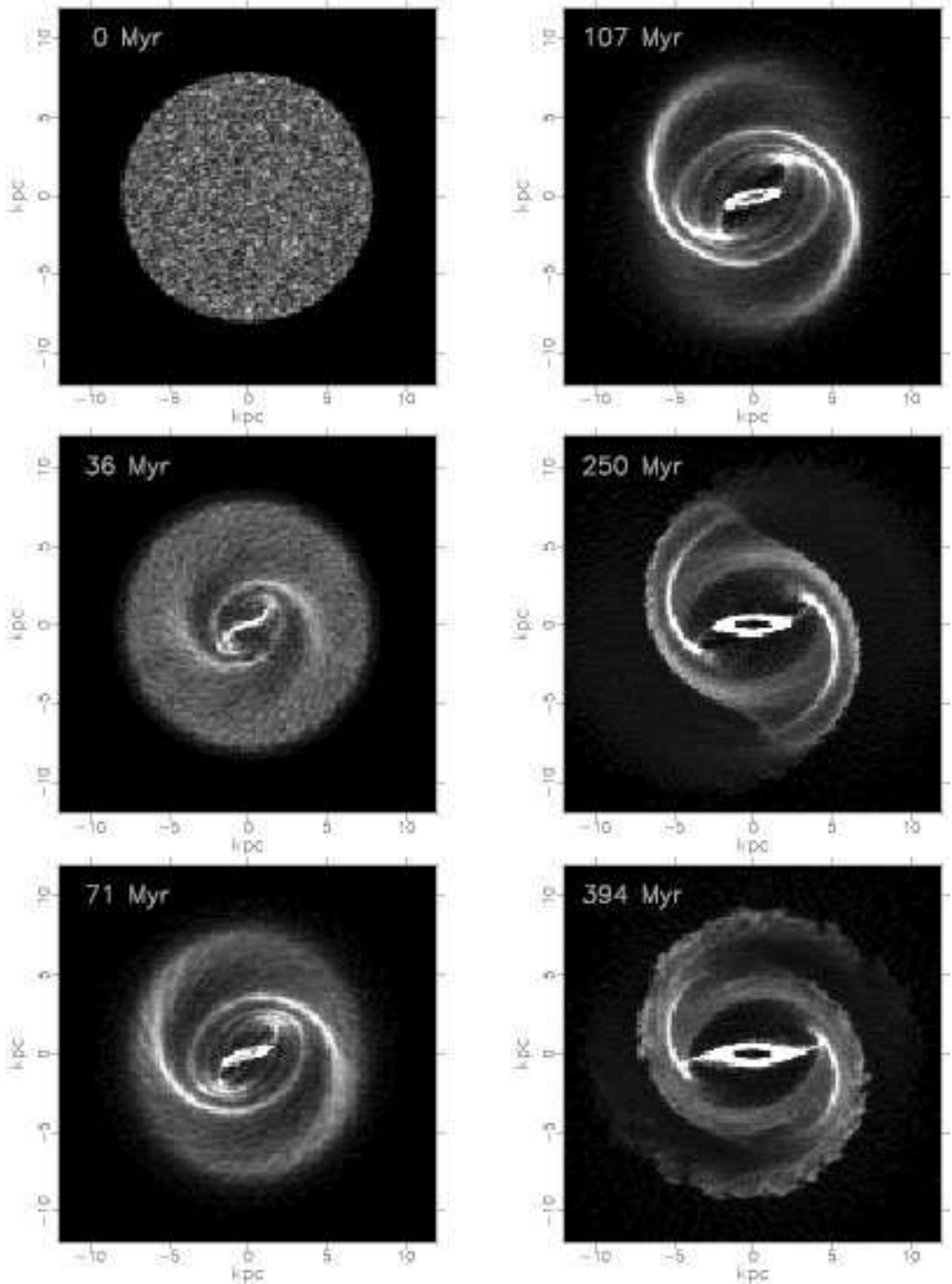}
      \caption{Gas dynamical evolution in model E:
($\omgb$, $\epsilon_0$) = (0.8, 0.10).
The stellar bar runs horizontally, and the gas rotates counterclockwise.
After $t\sim250\,{\rm Myr}$ the system reaches a state of
quasi-equilibrium.
              }
         \label{fig:evlv}
   \end{figure*}

Gas dynamics in a barred potential have been well studied in numerical simulations
\citep{wh92, hs94, ps95, fw98, ab99}. Our models evolve consistently with these
simulations. 
Fig. \ref{fig:evlv}, model E,  shows a typical evolution.
Three phases of the evolution can be seen in this model:
(a) linear perturbation phase, $t \sim 0-50\myr$,
(b) transient phase, $t \sim 50-250\myr$, and
(c) quasi-steady phase, $t>250\myr$.

The characteristic structure appearing during the evolution depends strongly
on the positions
of resonances, i.e. the pattern speed of the bar $\omgb$. In phase (a),
leading and trailing spiral arms are formed around the inner ($R=1.1\kpc$) and
outer ($3.1\kpc$) ILRs respectively at $t=36\myr$. 
These resonant-driven spirals are expected in a linear
theory \citep{wa94}.
While the outer trailing arms remain with increasing density contrasts,
the inner leading arms evolve into an oval ring, or a gaseous bar 
($t=71-107\myr$), i.e. phase (b).  The oval ring first leads the stellar bar
($71\myr$), rotating opposite to the gas rotation ($107\myr$), and being
aligned with the stellar bar ($250\myr$), and thereafter,  the system
develops toward a quasi-steady phase, i.e. phase (c). The ellipticity of
the nuclear ring grows as high as $e \sim 0.8$.
The ripple seen in the outer arms at $t=394\myr$ would originate
in the Kelvin-Helmholtz instability \citep{ps95}. 
Gas dynamics and structure in the inner region of the disk
are not affected by this instability.

%

   \begin{figure*}
   \centering
   \includegraphics[width=0.9\textwidth]{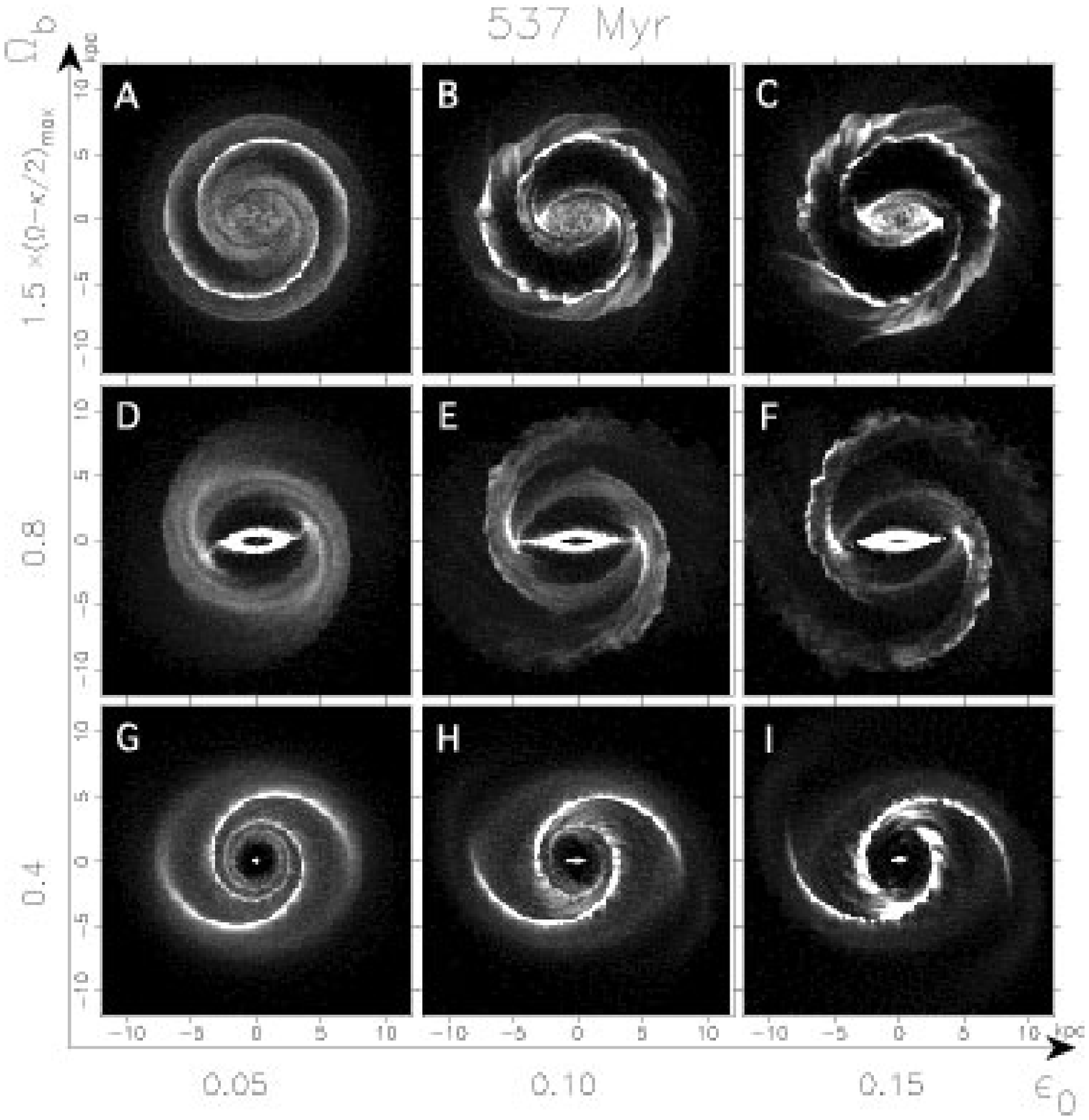}
      \caption{Final snapshots for all nine models. Different pattern speeds $\omgb$
    and bar strengths $\epsilon_0$ are arranged vertically and horizontally,
    respectively. The stellar bar runs horizontally, and the gas rotates
    counterclockwise.
              }
         \label{fig:fsht}
   \end{figure*}

%

   \begin{figure*}
   \centering
   \includegraphics[width=0.9\textwidth]{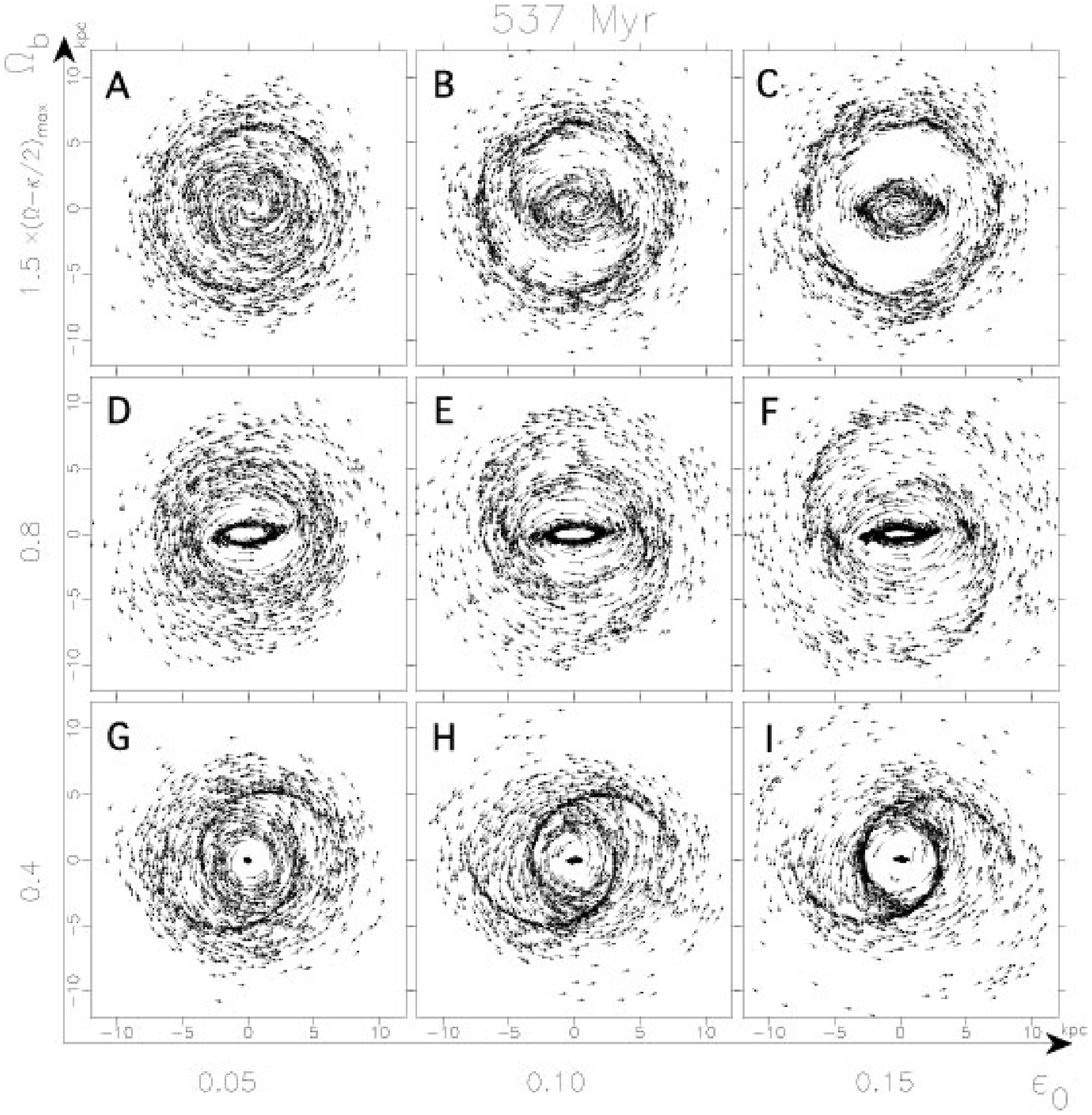}
      \caption{Final velocity fields for all nine models. The arrangement is
    the same as in Fig. \ref{fig:evlv}. Arrows are drawn for 1500 out of
    total 30000 gas particles.
              }
         \label{fig:velf}
   \end{figure*}


Fig. \ref{fig:fsht} and \ref{fig:velf} display the final snapshots and velocity
fields for nine models. Different pattern speeds $\omgb$ and bar strength
$\epsilon_0$ are arranged vertically and horizontally, respectively. It is evident
that the final structure depends strongly on $\omgb$,
 while $\epsilon_0$ changes
only the density contrasts.
Model A, B, and C have no ILRs, thus no spiral arms or ring in their inner
regions are formed.
The outer spiral arms are formed outside the radius of the corotation
resonance (CR) due to the outer Lindblad resonance (OLR).
Model D and F resemble model E. Model G, H and I also have arms and rings
similar to those in model E, but at different radii, corresponding to
the location of the ILRs.
Fig. \ref{fig:velf} clearly show that, in models D, E, F and G, most gaseous
orbits are $x_1$-like, while in model H and I, the large separation
and low density between the two ILRs suffice to leave the gases on
$x_2$-orbits, which form a stable oval ring, nearly perpendicular
to the stellar bar. 

\section{Galactic Mass Derived from Rotation Curves}
\label{sec:calprob}

In this section, we compare the apparent rotation curves 
obtained from the numerical results (Sect. \ref{sec:rotc}) and 
the true rotation curves, then we get probability to 
overestimate/underestimate the galactic mass (Sects. 3.2 and 3.3).

\subsection{Observing Rotation Curves in Models}\label{sec:rotc}

\begin{figure*}
   \centering
   \includegraphics[width=0.9\textwidth]{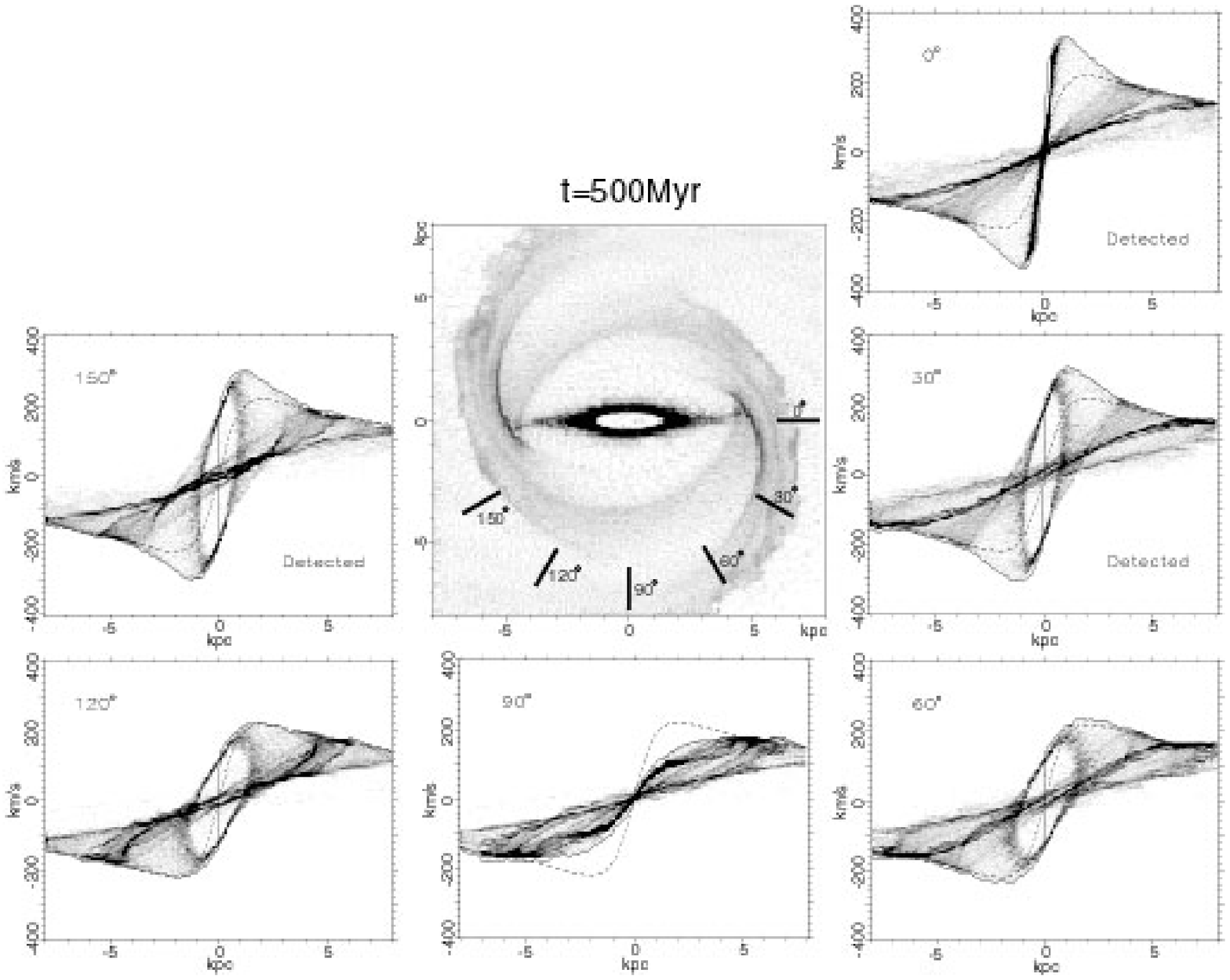}
   \caption{Position-velocity diagrams and rotation curves in model E,
    observed from a variety of viewing angles, at $t=500\myr$.
    Solid lines show the distorted rotation curves derived from
    the calculated p-v diagrams (gray scale), while dashed lines
    indicate the rotation curve derived from the model potential.
     The mark ``Detected'' indicates that our criterion with
    $\alpha_{\rm crit}=2$ is satisfied.}
              \label{fig:pvrc}%
    \end{figure*}

We obtain a position-velocity (p-v) diagram by observing our calculated gas
disks edge-on, and then determine a rotation curve from the p-v diagram.
We assign gas particles in a position-velocity grid using the cloud-in-cell
method \citep{he81}; the spacing for the grid is set to a typical resolution
in recent interferometry observations of the CO gas for Virgo galaxies,
i.e. $100\pc$ ($\sim1\arcsec$) in space and $5\kmps$ in velocity (Sofue in
private communication).
Then we determine a rotation curve by tracing the gas at the highest velocity
for each radius in the p-v diagram. Some examples for p-v diagrams and
rotation curves are shown in Fig. \ref{fig:pvrc}


There are several ways to derive a rotation curve
from an observed p-v diagram. One traces the peak-intensity velocity or
intensity-weighted mean velocity at each radius \citep{ru80,ru82,ru85,
mf92,mf96}, while another traces the 20\% envelope of the peak-intensity
velocity at each radius \citep{sf96}.
These intensity-based methods cannot be applied to our density-based
p-v diagram, because intensity is not a simple function of density,
especially in edge-on systems. 
Compared with these methods, our method provides generally higher
velocity.

\subsection{Error Estimation in Rotation Curves and Mass}\label{sec:over}

We compare the observed rotation curves $\vobsr$ derived from p-v diagrams
in simulations (Sect. \ref{sec:rotc}), with the true rotation curves $\vpotr$
from the gravitational potential ($\Phi_0(R)$). We estimate the errors in rotation curves by
defining a function, i.e.
\begin{equation}
\alpha(R) \equiv \frac{\vobsr}{\vpotr},
\label{eq:alpha}
\end{equation}
and examining whether  $\alpha$ exceeds an arbitrary critical value
$\alpcrit$ as
\begin{equation}
\alpha(R) \geq \alpcrit.
\label{eq:alpcrit}
\end{equation}
The ratio of the mass $\Mobs$, derived from an observed rotation curve
$\vobs$, over the true mass $\Mpot$ from the potential is,
\begin{equation}
\gamma(R) \equiv \frac{\Mobs}{\Mpot} [= \alpha(R)^2];
\label{eq:gamma}
\end{equation}
the second equal comes from Eq. (\ref{eq:vir}).
Then Eq. (\ref{eq:alpcrit}) becomes equivalent to
\begin{equation}
\gamma(R) \geq \gmmcrit,
\label{eq:gmmcrit}
\end{equation}
where $\gmmcrit = \alpcrit^2$.

A rotation curve rises steeply from a galactic center, having a peak, or
at least a shoulder, at an innermost region, then reaching the flat
rotation. A central galactic mass is always estimated at the radius of
the peak or shoulder in observations \citep[see][]{sf99}.
We thus consider the case that the mass
is overestimated/underestimated at the radius of the first peak or shoulder.
 The radius
depends on, and changes with a viewing angle for the gas disk (see Fig.
\ref{fig:pvrc}). Hence, we define a reference region of $400\pc < R < 2\kpc$
\footnote{The absolute values are not important because our results are
almost scalable for an arbitrary core radius.} to which the above criterion
, i.e. Eq. (\ref{eq:alpcrit}), is applied; in our
models the first peak or shoulder always fall in this region.
If Eq. (\ref{eq:alpcrit}) or (\ref{eq:gmmcrit}) is satisfied in this
reference region, our observed rotation velocity differs from
the true velocity at least by a factor of $\alpcrit$.

\subsection{Probability of Erroneous Mass Estimation}
\begin{figure}
   \centering
   \includegraphics[width=0.47 \textwidth]{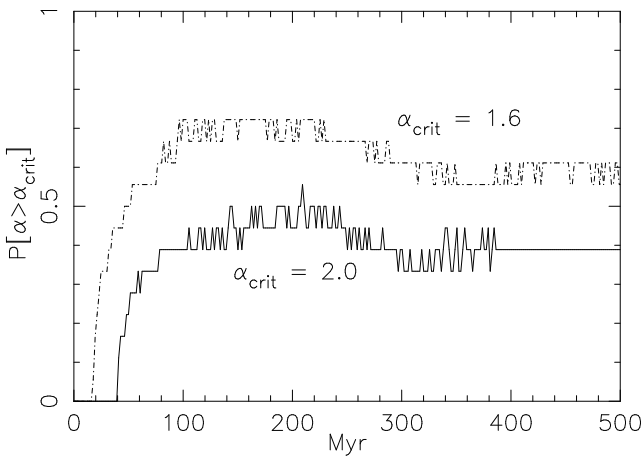}
   \caption{Change of the probability $P[\alpha>\alpcrit]=P_{\alpha}(\alpcrit)$
    that Eq. \ref{eq:alpcrit} will be satisfied. $\alpcrit = 1.6$ and
    $2.0$ indicate the cases that an observed rotation velocity exceeds
    the true value by a factor of1.6 and 2.0 times, respectively.}
              \label{fig:evlp}%
   \end{figure}

We define the probability that the central rotation velocity is
larger/smaller by a factor of $\alpcrit$, as the fraction of viewing
angles.
We obtain rotation curves by observing the gas disk for viewing angles
at ten-degree intervals, and calculate the probabilities
for three-hundred snapshots in each simulation. We hereafter describe
the probability as $P[\alpha>\alpcrit]$ or $P_{\alpha}(\alpcrit)$.
Fig. \ref{fig:evlp} shows a time evolution of $P_{\alpha}$ in model
E. 
Corresponding to the three evolutionary phases
(Sect. \ref{sec:gde}), the probability rises steeply in the linear
perturbation phase, approaching a constant value in the transient phase,
then reaching the constant in the quasi-steady phase. We also calculate
the averaged probability in the quasi-steady phase, i.e. $300-500\myr$,
and describe it as $\tilde{P}_{\alpha}$. 
Similarly $P_{\gamma}(\gmmcrit)$,
the probability that the estimated mass differs from the true galactic mass
by a factor of $\gmmcrit$, is defined. $\tilde{P}_{\gamma}$ is
a time-average of $P_{\gamma}(\gmmcrit)$.

%

   \begin{figure*}
   \centering
   \includegraphics[width=0.9 \textwidth]{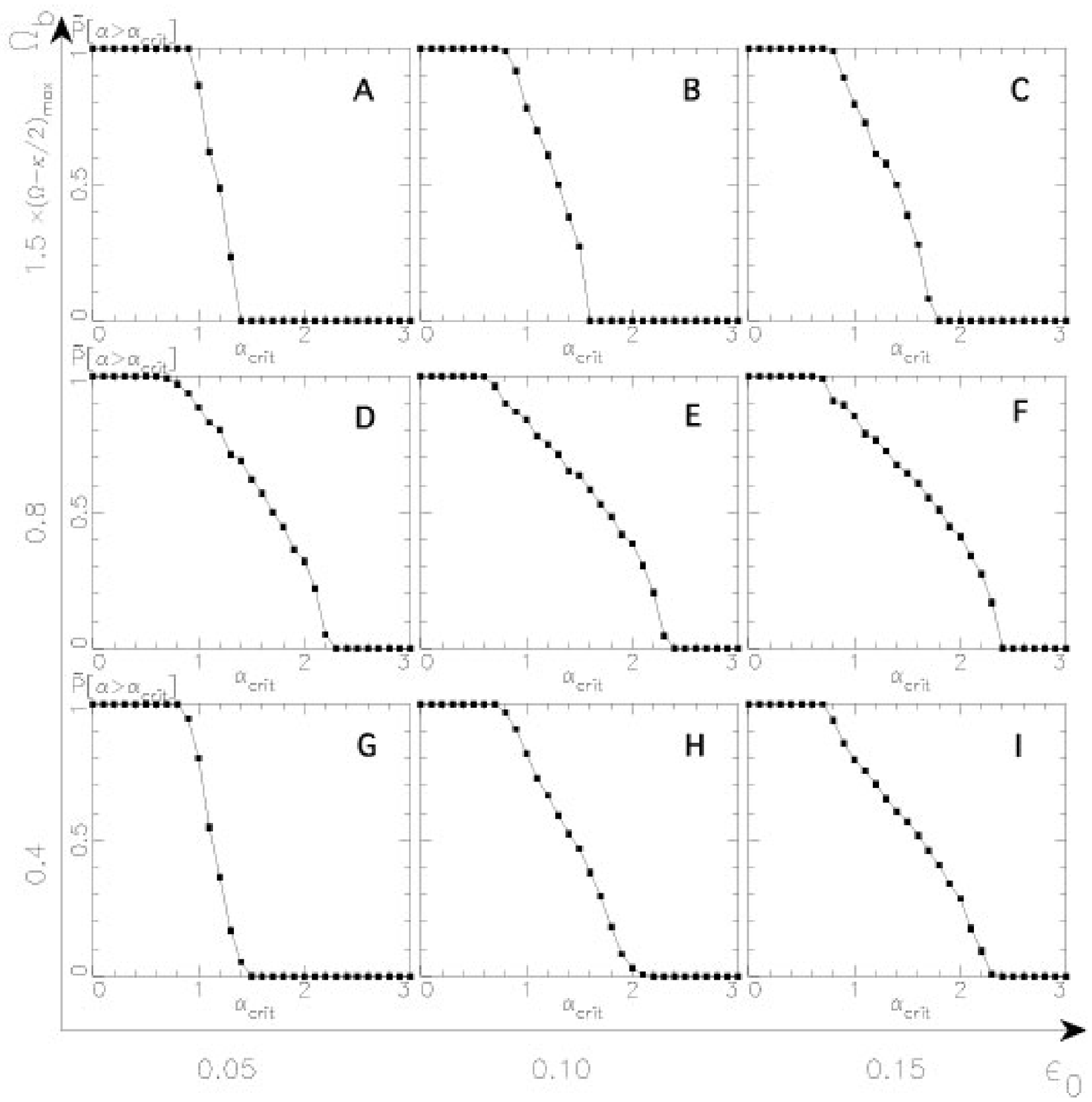}
   \caption{The probability that an observed rotation velocity would exceed
    $\alpcrit$ times that inferred from the galaxy potential.
    Probability $\avrPalp$ is averaged over the time $t=300-500\myr$ when
    the systems are in quasi-steady states.
    Models with different pattern speeds $\omgb$ and bar strengths
    $\epsilon_0$ are arranged vertically and horizontally, respectively.}
              \label{fig:res1}%
    \end{figure*}


   \begin{figure*}
   \centering
   \includegraphics[width=0.9 \textwidth]{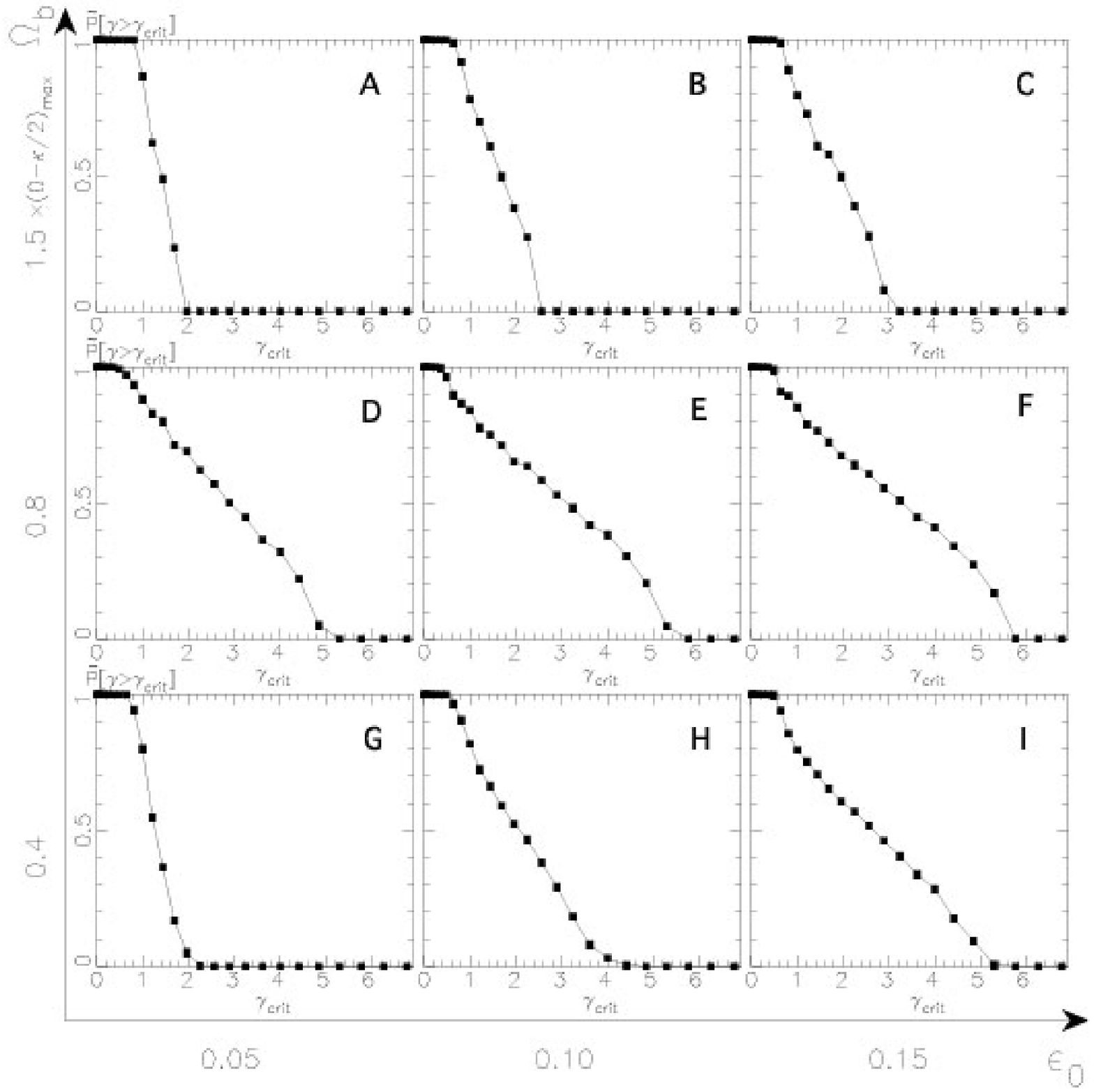}
   \caption{The probability that an observed mass would exceed
    $\gmmcrit$ times that inferred from the galaxy potential by chance.
    Probability $\avrPgmm$ is averaged in $t=300-500\myr$ when
    the systems are in quasi-steady states.
    Models with different pattern speeds $\omgb$ and bar strengths
    $\epsilon_0$ are arranged vertically and horizontally, respectively.}
              \label{fig:res2}%
    \end{figure*}


Fig. \ref{fig:res1} shows the averaged probability $\avrPalp$ as 
a function of  $\alpcrit$ for all nine models.
Different pattern speeds $\omgb$ and bar strengths $\epsilon_0$ are arranged
vertically and horizontally, respectively, as in Fig. \ref{fig:fsht}.
If the gas follows pure circular rotation, i.e. $\vobs = \vpot$, then these
$\avrPalp$-profiles must be a step function: $\avrPalp = 1$ for
$\alpcrit \leq 1$ and $\avrPalp = 0$ for $\alpcrit > 1$.
However, the non-circular motion changes the $\avrPalp$-profiles
as seen in the plots. In model E for example, $\avrPalp = 0.8$ at
$\alpcrit = 1.0$ means that 80\% of the rotation curves $\vobs$ observed
from random angles would apparently show higher velocities than the true
rotation curve $\vpot$ which traces the mass, and another 20\% would show
lower velocities than the true one. Thus the overestimation in mass
occurs more frequently than the underestimation in model E. $\avrPalp = 0.0$ at
$\alpcrit = 2.4$ indicates that the observed rotation curve cannot be
overestimated by more than the factor of 2.4 in model E.


We discussed in Sect. \ref{sec:gde} that the final structure depends strongly
on the pattern speed $\omgb$ and a little on the bar strength $\epsilon_0$.
This is also evident in Fig. \ref{fig:res1}; the global profiles in the
same $\omgb$ are quite similar, but $\avrPalp$ increases slightly with
increasing $\epsilon_0$.
Model A, B and C have no ILRs, not showing non-axisymmetric structures in
the central regions (Sect. \ref{sec:gde}), thus the $\avrPalp$-profiles
are similar to the step functions. Model D, E and F show the most prominent
streaming motions in their central regions, 
and therefore they have the largest $\avrPalp$.
 Although models G, H and I have the same $\omgb$,
the $\avrPalp$-profile for model G is different from those for models H
and I, because gaseous $x_2$-orbits remain in models H and I, but not in
model G (see Sect. \ref{sec:gde}).


Fig. \ref{fig:res2} shows the probability $\avrPgmm$ vs $\gmmcrit$.
All the plots show properties similar to those in Fig. \ref{fig:res1}.
In model E for example, $\avrPgmm=0.15$ at $\gmmcrit=5.0$ means that
the central galactic mass derived from an observed rotation curve is
overestimated by a factor of five in the probability of15\%.
$\avrPgmm$s
in all models become zero at $\gmmcrit=6.0$, meaning that the central
mass from an observed rotation curve can be overestimated by {\it at most}
a factor of six in our models.


\section{Discussion and Summary}\label{sec:disc}

Based on gas dynamical calculations in a fixed galactic potential with
a weak bar-like distortion,
 we estimated errors in mass estimation from the rotation curves, 
and calculated the probability that observations would
suffer from such errors. 
We found that, as well as the final morphologies of gas disks,
the probability strongly depends on the pattern speed of a bar $\omgb$,
and weakly on the bar strength $\epsilon_0$. Among our nine models,
the probability for the errors becomes maximal for the models
with $\omgb=0.8 \times (\Omega-\kappa/2)_{\rm max}$; the 15\%
of them have an error of a factor of five in mass estimation, if we observe 
the disks from an arbitrary viewing angle, and if we define
rotation curves as the highest-velocity envelope of the p-v diagrams.
Even in those erroneous cases, the galactic
mass is not overestimated by more than a factor of six.
In all of our models, the overestimation in mass is more probable
than the underestimation.
We consider only some particular cases for
a weak bar, thus cannot obtain general conclusions.
The above estimation however must be a guideline to consider the central
galactic mass derived from an observed rotation curve.



   \begin{figure*}
   \centering
   \includegraphics[width=0.9 \textwidth]{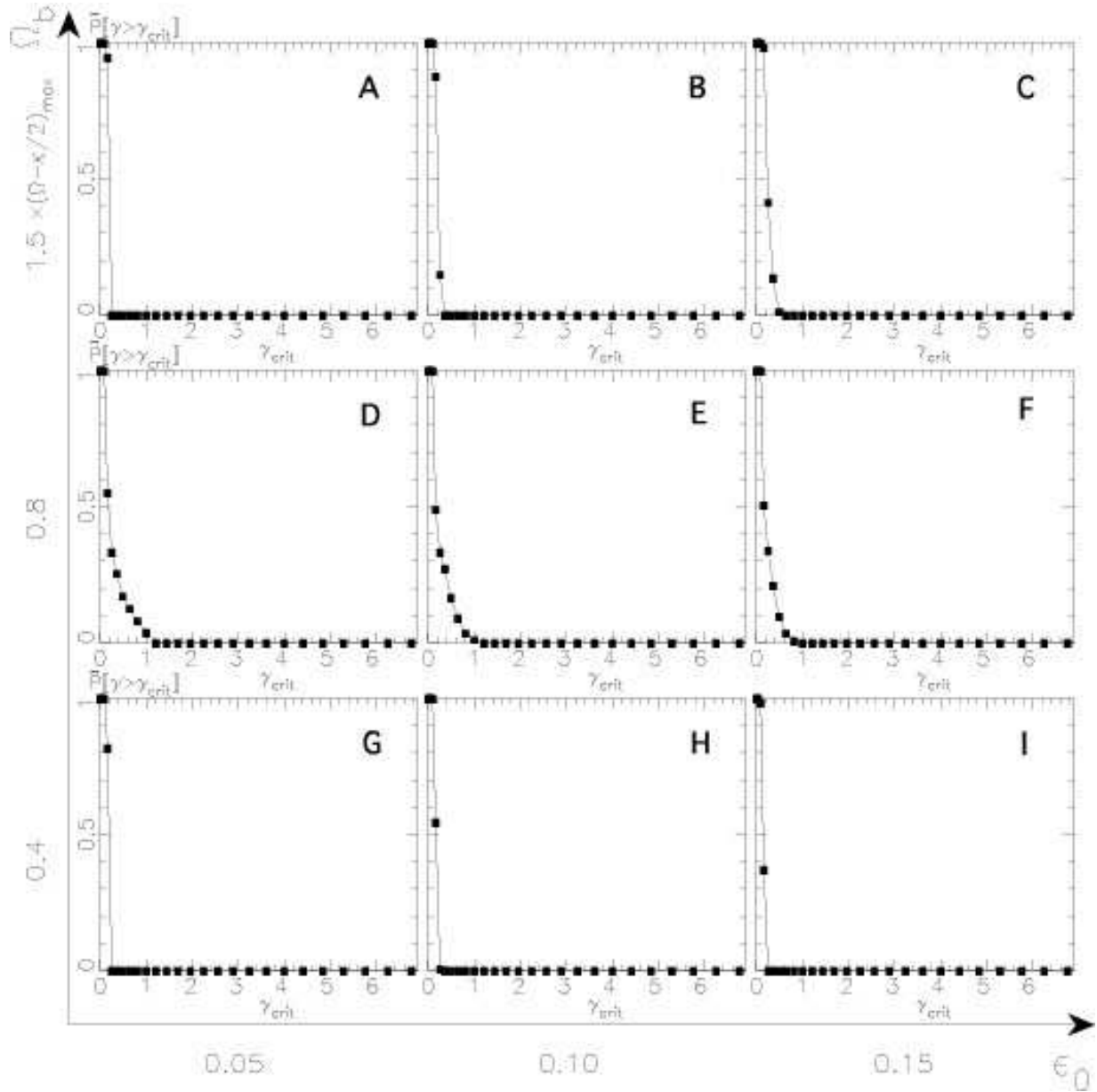}
   \caption{Same as Fig. \ref{fig:res2}, but rotation curves are derived
by taking the density-weighted mean velocity rather than the most rapidly
rotating envelope of the p-v diagram.
}
              \label{fig:res3}%
    \end{figure*}

Conventionally, rotation curves have been often defined as
the peak-intensity velocity or intensity-weighted mean velocity of
p-v diagrams. However, \citet{sf96} pointed out that these methods
underestimate the rotation velocity, particularly in the central region,
because the finite beam size causes the confusion with the gas with lower
line-of-sight velocities on the p-v diagram; this effect is also
demonstrated in \citet[][in their Fig. 15]{kd02}. For rotation curves
in highly inclined galaxies, this confusion can not be avoidable.
Alternatively, the envelope-velocity of the p-v diagram is better suited
to trace the central mass distribution \citep{sf96, sr01}.
Therefore we defined the highest-envelope velocity as our rotation curves
in the above study. Here we repeat the same analysis for a comparison,
using rotation curves derived from the density-weighted mean velocity,
and shows the results in Fig. \ref{fig:res3}.
$\avrPgmm$s are always less than those in Fig.
\ref{fig:res2}, and are almost zero at $\gmmcrit=1.0$. This means that
the mass derived from the mean-velocity rotation curves are
{\it almost always} underestimated in the central regions of galaxies.
These results suggest that the conventional method for deriving rotation
curves from p-v diagrams is not also relevant to estimate the mass in
galaxies with bar-like distortions.

\citet{sf99} showed that most of the rotation curves rise steeply from the
centers, reaching high velocities of about $100-300\kmps$ in the
innermost regions. Owing to the large fraction of the rotation curves with
these high central velocities, they discussed the idea
that these velocities should be attributed to massive cores rather than to
bars.
We may have a chance to statistically clarify whether or not the massive
cores exist by comparing a probability such as ours with the observed
fraction of rotation curves with high central velocities.
When we define the probability $P$ averaged in all types of barred
and non-barred galaxies by
\begin{equation}
P = \int \avrPalp(\omgb,\epsilon_0) f(\omgb, \epsilon_0) d\omgb d\epsilon_0,
\end{equation}
where $f(\omgb, \epsilon_0)$ is a distribution function of galaxies with
a pattern speed $\omgb$ and bar strength $\epsilon_0$, the existence of
massive cores
is confirmed if the fraction of galaxies with the central high velocities
is more than $P$. For example, using our maximum calculated probability
$\avrPalp|_{\rm max}=0.4$ and an observed fraction of barred galaxies
$f_{\rm bar}\sim0.6$ \citep{kn00}, $P$ could be very roughly calculated as
$P < \avrPalp|_{\rm max} \times f_{\rm bar} = 0.24$.
Of course, we need more intensive studies for
a number of barred potentials and parameters, and more precise knowledge
of the distribution function of parameters for bars.

\begin{acknowledgements}
We are grateful to Y. Sofue for fruitful discussions.
We also thank an anonymous referee and H. J. Habing,
the editor, for useful comments.
J. K. was financially supported by the Japan Society for the
Promotion of Science (JSPS) for Young Scientists.
\end{acknowledgements}

\end{document}